\documentclass[aps,pra,showkeys,twocolumn,superscriptaddress,floatfix]{revtex4-2}
\usepackage{array}
\usepackage{booktabs}
\usepackage{tabu}
\usepackage{dcolumn}
\usepackage{amsmath}
\usepackage{amsfonts}
\usepackage{float}
\usepackage{amssymb}
\usepackage{graphicx,color}
\usepackage[colorlinks={true}]{hyperref}
\hypersetup{colorlinks=true,linkcolor=red,citecolor=blue,urlcolor=blue}
\usepackage{graphicx}
\usepackage{subfigure}
\usepackage{graphicx}
\usepackage{dcolumn}
\usepackage{bm}
\usepackage{pstricks}
\usepackage{braket}
\usepackage{orcidlink}

\def\be{\begin{equation}}
  \def\ee{\end{equation}}
\def\bea{\begin{eqnarray}}
\def\eea{\end{eqnarray}}

\begin{document}
\title{Entanglement Swapping Using Hyperentangled Pairs of Two-Level Neutral Atoms}

\author{ Sajal Hasan}
\affiliation{Department of Physics, COMSATS University Islamabad, Islamabad, Pakistan}
\author{Syed M. Arslan \orcidlink{0000-0002-5820-6576}}\email{syan54988@hbku.edu.qa}
\affiliation{College of Science and Engineering, Qatar Center of Quantum Computing (QC2), Hamad bin Khalifa University, Doha, Qatar}
\author{Muhammad Imran} 
\affiliation{National Institute of Lasers and Optronics College
    Pakistan Institute of Engineering and Applied Sciences, Nilore, Islamabad, Pakistan}
\author{Rameez-ul Islam} 
\affiliation{National Institute of Lasers and Optronics College
    Pakistan Institute of Engineering and Applied Sciences, Nilore, Islamabad, Pakistan}
\author{Saif Al-Kuwari \orcidlink{0000-0002-4402-7710}}
\affiliation{College of Science and Engineering, Qatar Center of Quantum Computing (QC2), Hamad bin Khalifa University, Doha, Qatar}
\author{Tasawar Abbas\orcidlink{0000-0002-6048-7346}}
\affiliation{Department of Physics, COMSATS University Islamabad, Islamabad, Pakistan}

\title{Entanglement Swapping Using Hyperentangled Pairs of Two-Level Neutral Atoms}
\maketitle

\title{Entanglement Swapping Using Hyperentangled Pairs of Two-Level Neutral Atoms}


\maketitle
\section*{ABSTRACT}
Hyperentangled swapping is a quantum communication technique that involves the exchange of hyperentangled states, which are quantum states entangled in multiple degrees of freedom, to enable secure and efficient quantum information transfer. In this paper, we demonstrate schematics for the hyperentanglement swapping between separate pairs of neutral atoms through the mathematical framework of atomic Bragg diffraction, which is efficient and resistant to decoherence, yielding deterministic results with superior overall fidelity. The utilized cavities are in superposition state and interact with the incoming atoms off-resonantly. Quantum information carried by the cavities is swapped through resonant interactions with two-level auxiliary atoms. We also discuss entanglement swapping under a delayed-choice scenario and provide a schematic generalization covering multiple-qubit scenarios. Finally, we introduce specific experimental parameters to demonstrate the experimental feasibility of the scheme.       

    \textbf{Keywords}: Quantum Communication, Quantum Information, Quantum Optics
\section{Introduction}
Entanglement is an essential component in many quantum information technologies, including quantum computing \cite{gupta2020quantum}, quantum sensing\cite{pirandola2018advances, zhuang2018distributed}, and, most notably, quantum communication \cite{giddings2018quantum,zidan2018novel,sagheer2019novel,shor1999polynomial}.

Many quantum communication protocols based on entanglement have been developed and experimentally demonstrated \cite{haq2015long,pan1998experimental,bose1998multiparticle,ma2012experimental,goebel2008multistage,ali2022hyperentanglement, ali2022teleportation}. One of these protocols is Entanglement Swapping, which is a quantum phenomenon that allows the entanglement of two particles, even if they have never directly interacted. In fact, it has many applications in quantum information\cite{haq2015long,pan1998experimental,bose1998multiparticle}. Recently, many quantum communication protocols began to use a variant of the normal quantum entanglement, known as Hyperentanglement, where the entanglement takes place between multiple degrees of freedom.
Hyperentangled states improve channel capacity by minimizing the physical resources, and hence, in certain situations, lower the decoherence linked to the number of quantum entities involved \cite{islam2008engineering,sun2023complete,zeng2020complete}. The swapping of such hyperentangled states presents a viable solution for efficient quantum information transfer that requires minimal physical resources, especially in the context of information distribution over complex quantum networks \cite{anis2021engineering,xiong2016multiple}.  

In this paper, we engineer hyperentanglement with its internal degree of freedom being molded by the atom's ability to be in the ground or in the excited state, and use the transverse quantized momentum state of the atom as its external degree of freedom. We use Atomic Bragg diffraction to generate hyperentangled atomic states i.e. states entangled simultaneously in atomic internal and external degrees of freedom and their further manipulation for intended entanglement swapping. Atomic Bragg diffraction is one of the most exciting applications of cavity quantum electrodynamics, which is often used for quantum state engineering and atomic interferometry and can be used to generate hyperentanglement with external and internal degrees of freedom of atoms \cite{ul2013extended,anis2021engineering}. Atomic Bragg diffraction is a technique through which one can develop different types of optical gadgets such as atomic beam splitters and atomic mirrors \cite{haroche2013nobel,ikram2019entanglement,imran2021quantum,qureshi2023exploring}.

Our scheme involves three parties: Alice, Bob, and Carol. First, Alice and Bob mutually generate a hyperentangled pair of two-level atoms. Engineering of such states involves off-resonant atomic Bragg diffractions with the cavity field, which is initially in a superposition state of zero and one followed by exposing a momentum arm to a classical laser. Atoms are initially in the ground state $\ket{g}$ with the momentum state $\ket{P_0}$. The interactions are first-order Bragg diffractions. Consequently, after controlled exposure to the laser field, the hyperentangled atomic state is produced, where the internal and external (transverse-momentum) degrees of freedom are correlated. Once Alice and Bob generate such state, each one sends 1 qubit (atom) to a third-party, Carol, who possesses two further cavity fields in superposition. Carol adjusts the cavities so that the deflected arm of each atom passes through one cavity, and the undeflected arm of each atom passes through another cavity. Finally, Alice and Bob detect the internal and external states of the two atoms through four different state-selective detectors after passing them through a Ramsey zone. Therefore, this results in the remaining two atoms, which never interacted with each other, sharing an entangled state. This completes the entanglement swapping process, in which the entanglement is swapped between the entangled pairs that were initially prepared.

The rest of this paper is organized as follows: In section \ref{ABD}, review the mathematical framework of atomic Bragg diffraction along with the generation of hyperentangled Bell state among couple of two-level atoms. In Section \ref{Ent_swap}, we describe our entanglement swapping scheme. Section \ref{delayed} describes the delayed-choice hyperentangled swapping along with its due generalization. Section \ref{experiment} briefly elaborates on the experimental feasibility envisioned for the proposal under the prevailing experimental scenario. Finally, Section \ref{conclusion} concludes the paper.

\section{Atomic Bragg Diffraction}
\label{ABD}
Atomic Bragg diffraction is a Raman scattering phenomenon that follows the general conditions of Bragg diffraction. Thus, the conservation of momentum and energy of the system remains intact. Momentum exchange occurs between the quantized field trapped in a cavity and the atom, where the magnitude remains constant, and only the direction of the external momenta changes. The atom's momentum consists of two components: transverse and longitudinal. The atom travels at a slight angle relative to the axis of the cavity, which results in the quantization of the transverse component of the momentum, and the longitudinal component, still large, may be treated classically. When such an atom interacts with the quantized cavity field, the atom experiences a momentum kick and, consequently, discrete multiples of quantized atomic momentum are transferred along the k-vector of the field. This is provided by the relation $ P_{out}=P_{in} + l\hbar k $, where $P_{in}=P_0=l_0\hbar k \big/ 2$ is the initial momentum of the atom and $l_0$ is an even integer that refers to the order of Bragg diffraction. Each interaction transfers a momentum kick to the atom of magnitude 0 or $2\hbar k$ and thus $P_{out}$ is the resultant momentum after $l$ such interactions \cite{krahmer1994atom}. Furthermore, the relation of energy conservation is $|p_{in}|^2 \big/2M = |p_{out}|^2 \big/2M$, where $M$ is the mass of the atom. Using the relations for conservation of momentum and energy, we obtain the fundamental principle of atomic Bragg diffraction: $l(l+l_0)\hbar^2 k^2 / 2M=0$, which consists of two solutions, the first is $l=0$ representing the undeflected momentum state and the second is $l=-l_0$ representing the deflected momentum state. Similarly, first, second and third-order Bragg diffractions will be represented by higher-order of even integers 2, 4 and 6, respectively \cite{khan1998quantum, khan1999quantum}. 

\begin{figure}[t]
      \centering
    \includegraphics[width=8cm]{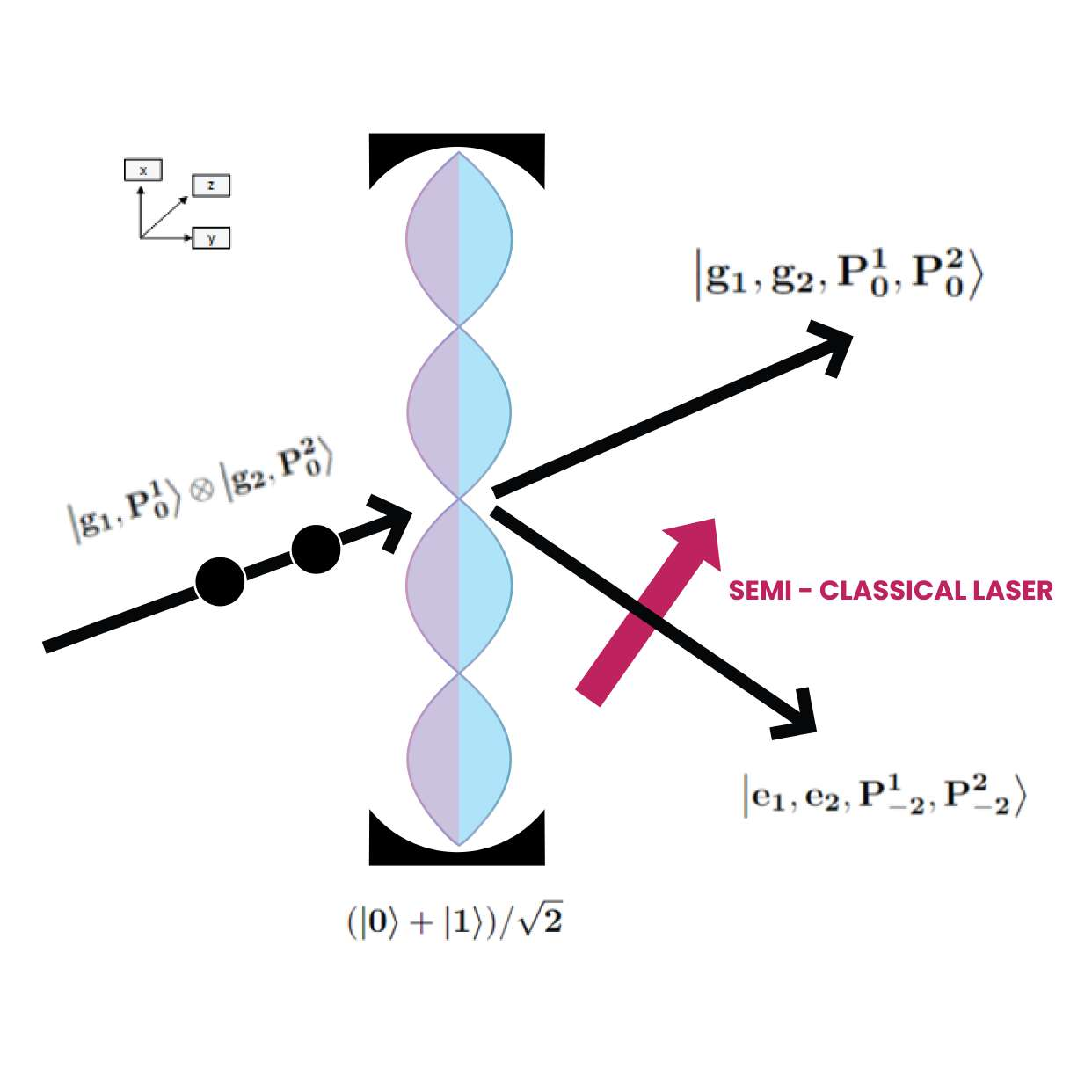}
    \caption{Schematics of the hyperentangled Bell state generation using off-resonant Bragg diffraction of two atoms initially in states $\ket{g_1,P_0^1}$ and $\ket{g_2,P_0^2}$ injected sequentially in a cavity which is in superposition state $(\ket{0}+\ket{1})/\sqrt{2}$.}
    \label{fig1}
\end{figure}

To demonstrate the phenomena of first-order Bragg diffraction, consider a two-level atom in its ground state $\ket{g}$ and quantized external momentum state $\ket{P_0 = \hbar k}$ on the x-axis, i.e., cavity axis, with the atom moving in the z-axis direction with longitudinal classical momentum, as shown in Fig. (\ref{fig1}). Meanwhile, the cavity is initially taken to be in superposition state $(\ket{0}+\ket{1})/\sqrt{2}$. The interaction picture Hamiltonian under dipole and rotating wave approximation is \cite{khan1998quantum, khan1999quantum,nawaz2017engineering,nawaz2018remote,nawaz2019atomic,abbas2014biasing,ikram2015wheeler,islam2013generation,saif2009engineering,ikram2007engineering,islam2008engineering,ul2009atomic,khosa2004measurement,khalique2003engineering,ikram2015wheeler},
\begin{equation} \label{Eq:1}
    H_I=\frac{P_x^2}{2M}+ \frac{\hbar \Delta}{2} \sigma_z+ \hbar \mu \cos(kx) \left[b\sigma_+ + \sigma_- b^\dagger \right].
\end{equation} 
Here, the term $P_x^2/2M$ represents the kinetic energy of the atom associated with the quantized momentum component, and $\sigma_-$, $\sigma_+$ and $\sigma_z$ are the atomic lowering, raising, and inversion operators, respectively. $\Delta$ represents atom-field de-tuning and $\mu$ is the coupling constant between the atom and field with $b$ and $b^\dagger$ being the annihilation and creation field ladder operators. 
In order to build a correlation between the internal and external degrees of freedom of a pair of atoms, we begin with consideration of an atom in initial ground state $\ket{g}$ with quantized momentum state $\ket{P_0}$ which interacts with a cavity field in a superposition state $(\ket{0}+\ket{1})/\sqrt{2}$ as represented in Fig. (\ref{fig1}). The initial state vector is $\ket{\psi(t=0)} = \ket{g,P_0} \otimes (\ket{0}+\ket{1})/\sqrt{2}$. This specific initial state can be used to engineer the hyper-superposition state of a single atom \cite{haq2015long,anis2021engineering}. For this the proposed state vector can be written as,
\begin{equation} \label{Eq:2}
\begin{split}
    \left| \psi (t) \right> = & e^{-\Dot{\Dot{\iota}} \left( \frac{P_0^2}{2M \hbar}-\frac{\Delta}{2}\right)t} \sum_{l=-\infty}^{\infty} \left[ C_{g,0}^{P_l}(t)\left|g,0,P_l\right> \right.\\&\left.+  C_{g,1}^{P_l}(t) \left|g,1,P_l\right> +  C_{e,0}^{P_l}(t) \left|e,0,P_l\right> \right],
\end{split}    
\end{equation}
where $C_{g,0}^{P_l}(t), C_{g,1}^{P_l}(t)$ and $C_{g,1}^{P_l}(t)$ are the unknown time-evolved probability amplitudes of the respective states. It should be noted that $C_{g,0}^{P_l}(t)$ represents the possibility that the atom remains in the ground state $\ket{g}$ leaving the cavity in fock state $\ket{0}$, while $C_{g,1}^{P_l}(t)$ represents the possibility that the atom leaves the cavity in fock state $\ket{1}$ and still remains in ground state $\ket{g}$ and finally $C_{g,1}^{P_l}(t)$ represents the case where the internal state of the atom gets excited $\ket{e}$ leaving the cavity in fock state $\ket{0}$, after $l$ interactions. Solving the Schrodinger equation using the proposed state vector given in Eq. (\ref{Eq:2}) with the Hamiltonian expressed in Eq. (\ref{Eq:1}), we obtain the following differential equations for the rate of change of the probability amplitudes. We note that expression (\ref{Eq:3}) is a simple first-order differential equation, while Eq. (\ref{Eq:4}) and (\ref{Eq:5}) represent an infinite set of coupled differential equations,
\begin{equation} \label{Eq:3}
        \Dot{\Dot{\iota}} \frac{\partial}{\partial t} C_{g,0}^{P_l}(t) = \left( \frac{l(l_0 +l)\hbar k^2}{2M} \right) C_{g,0}^{P_l}(t),
\end{equation}
\begin{equation} \label{Eq:4}
\begin{split}
    \Dot{\Dot{\iota}} \frac{\partial}{\partial t} C_{g,1}^{P_{l}}(t) =& \left( \frac{l(l_0 +l)\hbar k^2}{2M} \right) C_{g,1}^{P_{l}}(t) \\& + \frac{\mu}{2} \left( C_{e,0}^{P_{l+1}}(t) + C_{e,0}^{P_{l-1}}(t) \right),
\end{split}
\end{equation}
\begin{equation} \label{Eq:5}
\begin{split}
    \Dot{\Dot{\iota}} \frac{\partial}{\partial t} C_{e,0}^{P_l}(t) = &\Delta C_{e,0}^{P_l}(t) \\&+ \frac{\mu}{2} \left[ C_{g,1}^{P_{l+1}}(t) + C_{g,1}^{P_{l-1}}(t) \right].
\end{split}    
\end{equation}
In this case, the atom-cavity interaction is considered to be off-resonant and obeys the Bragg diffraction condition, i.e $\Delta>\omega_r$ where $\omega_r=\hbar k^2/2M$ is the recoil frequency imparted to the atom. The adiabatic assumption is applied where the contribution of the slowly varying probability amplitudes is considered negligible \cite{khalique2003engineering}. Initially, the atom is in the ground state $\ket{g}$, therefore, Eq. (\ref{Eq:4}) shows an even number of interactions and Eq. (\ref{Eq:5}) shows an odd number of interactions. Setting $l_0 = 2$ for first order Bragg diffraction, we get, \\\\
For l=1, \begin{equation}
\begin{split}
\Dot{\Dot{\iota}} \frac{\partial}{\partial t}C_{e,0}^{P_{1}}(t)= \Delta C_{e,0}^{P_{1}} (t) + \frac{\mu}{2} \left(C_{g,1}^{P_{2}}(t) + C_{g,1}^{P_{0}}(t) \right),
\end{split}
\end{equation} 
\\
For l=0, \begin{equation}
\begin{split}
    \Dot{\Dot{\iota}} \frac{\partial}{\partial t}C_{g,1}^{P_0}(t) =
    \frac{\mu}{2} \left( C_{e,0}^{P_{1}} (t) +C_{e,0}^{P_{-1}} (t) \right),
\end{split}
\end{equation} 
\\
For l=-1,
\begin{equation}
\begin{split}
    \Dot{\Dot{\iota}} \frac{\partial}{\partial t}C_{e,0}^{P_{-1}}(t)= \Delta C_{e,0}^{P_{-1}} (t) + \frac{\mu}{2} \left(C_{g,1}^{P_{0}}(t) + C_{g,1}^{P_{-2}}(t) \right), 
\end{split}
\end{equation} 
\\
For l=-2,
\begin{equation}
\begin{split}
     \Dot{\Dot{\iota}} \frac{\partial}{\partial t}C_{g,1}^{P_-2}(t) = \frac{\mu}{2} \left( C_{e,0}^{P_{-1}} (t) +C_{e,0}^{P_{-3}} (t) \right),
\end{split}
\end{equation} 
\\
For l=-3,
\begin{equation}
\begin{split}
    \Dot{\Dot{\iota}} \frac{\partial}{\partial t}C_{e,0}^{P_{-3}}(t)= \Delta C_{e,0}^{P_{-3}} (t) + \frac{\mu}{2} \left(C_{g,1}^{P_{-2}}(t) + C_{g,1}^{P_{-4}}(t) \right).
\end{split}
\end{equation} 

Furthermore, under the adiabatic approximation, the time evolution of the probability amplitudes $\partial C_{g,1}^{P_{1}}(t)/\partial t, \partial C_{g,1}^{P_{-1}}(t)/\partial t$ and $\partial C_{g,1}^{P_{-3}}(t)\partial t$ can be ignored on practical grounds and the higher order amplitudes $C_{g,1}^{P_{2}}(t)$ and $C_{g,1}^{P_{-4}}(t)$ may also be skipped due to meager contributions \cite{ikram2007engineering,islam2008engineering,islam2008generation,khosa2004measurement,khalique2003engineering,ikram2015wheeler}. Under these approximations, the above infinite set of coupled equations reduces to the following two coupled differential equations:
\begin{equation}
    \frac{\partial}{\partial t} C_{g,1}^{P_0}(t) =\Dot{\Dot{\iota}} \beta t \left[ 2C_{g,1}^{P_0}(t) + C_{g,1}^{P_-2}(t) \right],
\end{equation} 
\begin{equation}
   \frac{\partial}{\partial t}  C_{g,1}^{P_{-2}}(t) = \Dot{\Dot{\iota}} \beta t \left[ C_{g,1}^{P_-2}(t) +  C_{g,1}^{P_0}(t) \right],
\end{equation}
with $\beta=\mu^2/4\Delta$. By solving these equations, we retrieve the time-evolved probability amplitudes associated with the state vector given in Eq. (\ref{Eq:2}):
\begin{equation}  \label{Eq:13}
    \begin{split}
    C_{g,1}^{P_0}(t) =& e^{2\Dot{\Dot{\iota}} \beta t} \left[  C_{g,1}^{P_0}(0) \cos\left(\beta t\right)  + \Dot{\Dot{\iota}} C_{g,1}^{P_-2}(0)\sin\left(\beta t\right) \right],
    \end{split}
\end{equation}
\begin{equation} \label{Eq:14}
    \begin{split}
    C_{g,1}^{P_{-2}}(t) =& e^{2\Dot{\Dot{\iota}} \beta t} \left[  C_{g,1}^{P_-2}(0) \cos\left(\beta t\right) + \Dot{\Dot{\iota}} C_{g_1,1}^{P^1_0}(0)\sin\left(\beta t\right) \right].
    \end{split}
\end{equation} 
By selecting the appropriate initial probability amplitudes, one can engineer the hyper-superposition state. In this specific case, we have selected $C_{g,1}^{P_0}(0) =C_{g,0}^{P_0}(0)=1/\sqrt{2}$ and $C_{g,1}^{P_{-2}}(0)=0$ together with the appropriate interaction time as $t= 2 \pi \Delta/\mu^2$ we get the state \cite{ul2013extended,anis2021engineering}:
\begin{equation} \label{Eq:15}
    \ket{\psi(t=2 \pi \Delta/\mu^2)} = \frac{1}{\sqrt{2}} \left[ \left| 0,g_1,P_{0} \right> -\Dot{\Dot{\iota}} \left| 1,g_1,P_{-2} \right> \right].
\end{equation} 
However, if initially the atom would have been considered to be in excited state, the same mathematical procedure would be followed, which is presented in Appendix A.

To generate the hyper-superposition state between the internal and external degrees of freedom of a single atom, the lower deflected momenta arm i.e. $\ket{P_{-2}}$ is exposed to the classical laser beam along the y-axis, as depicted in Fig. (\ref{fig1}). This semi-classical  resonant interaction \cite{zubairy1997quantum} is governed by the Hamiltonian, $(\hbar \Omega/2)[e^{-\Dot{\Dot{\iota}} \phi}\sigma_+ + e^{\Dot{\Dot{\iota}} \phi}\sigma_-]$, where $\Omega$ is the Rabi frequency of the atom and choosing the interaction time as $t=\pi/\Omega$, the state becomes:
\begin{equation} \label{Eq:16}
    \ket{\psi(t=\pi/\Omega)} = \frac{1}{\sqrt{2}} \left[ \left| 0,g_1,P^1_{0} \right> -\Dot{\Dot{\iota}} \left| 1,e_1,P^1_{-2} \right> \right].
\end{equation}
Here, it is apparent that a correlation has been built between the atom's external and internal states, and the cavity field. The subscripts of the internal states and the superscripts of the external states represent the first atom to pass through the cavity field. Passing a second atom, initially in state $\ket{g_2,P_0^2}$ from the same atom-cavity system, a correlation will be generated between the external states of both atoms with the cavity field. This second atom will follow similar conditions \cite{ul2013extended,anis2021engineering} to off-resonant Bragg diffraction as described earlier, and following the same mathematical procedure, which will obtain the probability amplitudes as described in the equation. (\ref{Eq:13}) and Eq. (\ref{Eq:14}). This second atom will again pass through the semi-classical laser field obeying resonant interaction and the state of the system will become:
\begin{equation} \label{Eq:17}
    \left| \psi(t) \right> = \frac{1}{\sqrt{2}} \left( \left|0, g_1,g_2,P_{0}^1,P_{0}^2 \right> - \left|1, e_1,e_2,P_{-2}^1,P_{-2}^2 \right> \right).
\end{equation} 
When the cavity of the atomic system is traced out, a hyperentangled bipartite Bell state \cite{ul2013extended} will be generated. For this, an auxiliary two-level atom initially in the ground state interacts resonantly with the cavity field. This resonant interaction is governed by the Hamiltonian, $H_R = \hbar \mu \left[ \sigma_+ b + \sigma_- b^{\dagger} \right]$, with $b(b^{\dagger})$ being the ladder operators for the field and $\sigma_+(\sigma_-)$ are transition operators for the atom. This resonant auxiliary atom leaves the cavity field into the vacuum state and is detected in either of its internal states after passing through the Ramsey zone, which allows us to engineer two mutually orthogonal equi-probable states. For brevity, we take the following engineered state:
\begin{equation} \label{Eq:18}
    \left| \psi(t) \right> = \frac{1}{\sqrt{2}} \left( \left| g_1,g_2,P_{0}^1,P_{0}^2 \right> -\Dot{\Dot{\iota}} \left| e_1,e_2,P_{-2}^1,P_{-2}^2 \right> \right).
\end{equation} 
The above equation represents the hyperentangled Bell state, which shows that the internal and external degrees of freedom of both atoms are entangled \cite{nawaz2017engineering}. This basic hyperentangled state acts as a building block for our proposal of hyper-entanglement swapping.

\section{Entanglement Swapping}
\label{Ent_swap}
\begin{figure}[t]
       \centering
     \includegraphics[width=9.5cm]{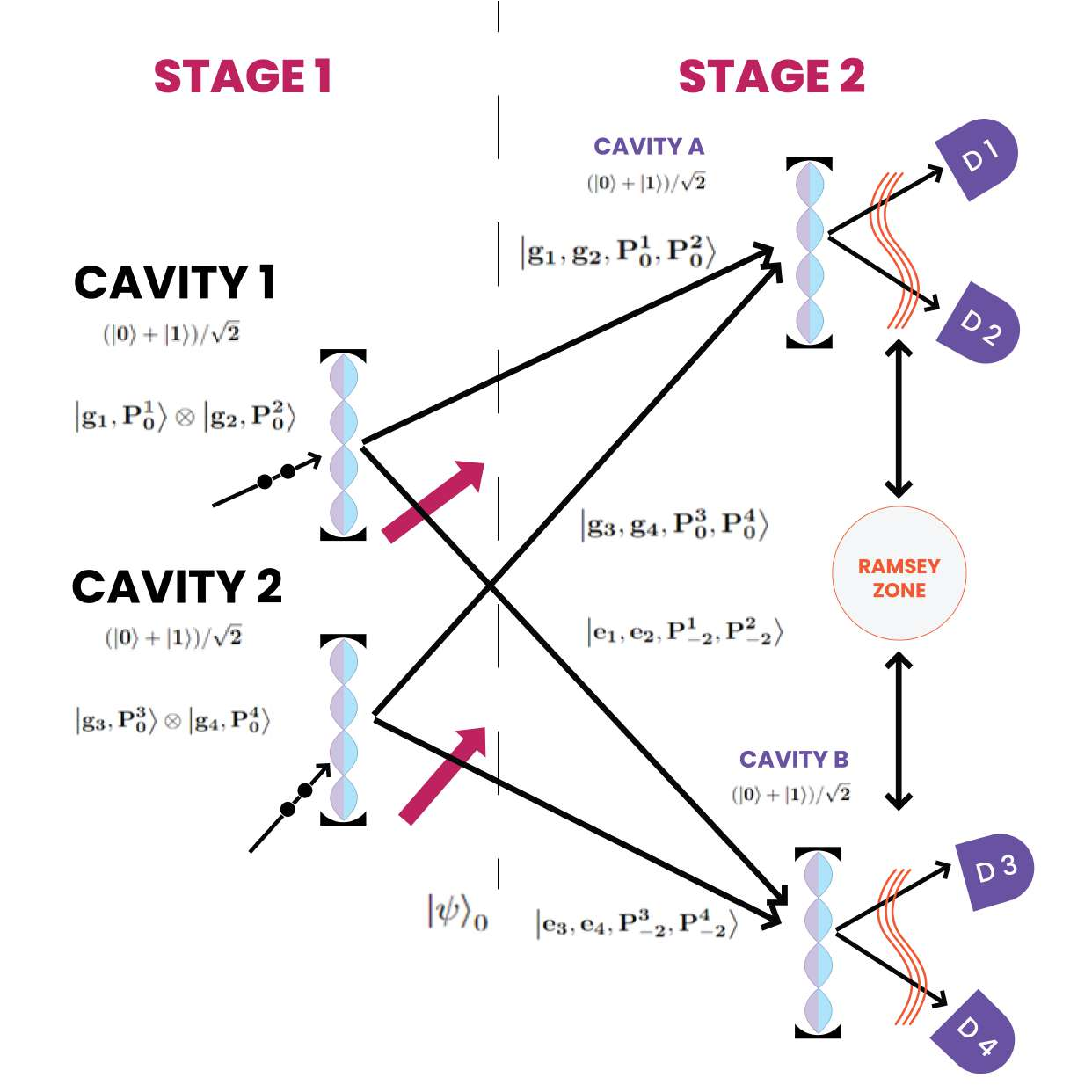}
     \caption{Illustration for implementing Quantum Swapping.}
     \label{fig2}
\end{figure}

Fig. (\ref{fig2}) visually describes our proposed scheme, which consists of two stages: 1) the generation of a bipartite hyperentangled state, 2)  swapping to another bipartite hyperentangled state. At stage 1 in Fig. (\ref{fig2}), two independent parties, Alice and Bob, generate hyperentangled pairs of atoms 1 and 2, and atoms 3 and 4, respectively, as described by Eq. \ref{Eq:18}. Once these pairs are generated, the initial cavities (Cavity 1 and 2) are traced out. This produces two pairs of hyperentangled atoms, which can be jointly written as: 
\begin{align} \label{Eq:19}
\ket{\psi_{0}} =  \frac{1}{2} \big[ \big( \ket{g_1,g_2,P_0^1,P_0^2}-\Dot{\Dot{\iota}}\ket{e_1,e_2,P_{-2}^1,P_{-2}^2}\big) \notag
 \\ \otimes \big( \ket{g_3,g_4,P_0^3,P_0^4}-\Dot{\Dot{\iota}}\ket{e_3,e_4,P_{-2}^3,P_{-2}^4}\big) \big].
\end{align}
Eq. \ref{Eq:19} show that there is no correlation between atoms 1 and 4 and similarly between atoms 2 and 3. In stage 2, we swap this hyper-entanglement on the pair of atom 1 and atom 4 by performing Bell basis measurement of the internal and external states of atoms 2 and 3. We consider two cavities (A and B) that exist in the superposition state $(\ket{0}+\ket{1})/\sqrt{2}$. Then, the joint state of the two pairs of hyperentangled atoms, together with the cavities, can be written as:
\begin{equation} \label{Eq:20}
    \begin{split}
        \left| \psi \right> = & \frac{1}{2} \left[ \left| g_2,P_0^2\right> \left|g_3,P_0^3\right>\left|g_1,g_4,P_0^1,P_0^4\right> \right.\\&\left. -\Dot{\Dot{\iota}} \left| g_2,P_0^2\right> \left|e_3,P_{-2}^3\right>\left|g_1,e_4,P_0^1,P_{-2}^4\right> \right. \\& \left.  -\Dot{\Dot{\iota}} \left| e_2,P_{-2}^2\right> \left|g_3,P_0^3\right>\left|e_1,g_4,P_{-2}^1,P_0^4\right> \right.\\&\left. - \left| e_2,P_{-2}^2\right> \left|e_3,P_{-2}^3\right>\left|e_1,e_4,P_{-2}^1,P_{-2}^4\right> \right] \\& \otimes \frac{1}{\sqrt{2}} \left(\left|0_A\right>+\left|1_A\right>\right)  \otimes \frac{1}{\sqrt{2}} \left(\left|0_B\right>+\left|1_B\right>\right).
    \end{split}
\end{equation}
The undeflected atomic wavepackets of atom 2 and atom 3, namely $\ket{g_1,P_0^2}$ and $\ket{g_3,P_0^3}$, respectively, interact with cavity A. Similarly, the deflected atomic wavepacket, namely $\ket{e_1,P_{-2}^2}$ and $\ket{e_3,P_{-2}^3}$, respectively, interact with cavity B. This is illustrated in Fig. (\ref{fig2}). Hence, Eq. \ref{Eq:20} can be re-ordered as:
\begin{equation}
    \begin{split}
        \ket{\psi} = & \frac{1}{2} \Big[\ket{g_2,P_0^2}\ket{g_3,P_0^3}\ket{A} -\Dot{\Dot{\iota}} \ket{g_2,P_0^2}\ket{e_3,P_{-2}^3}\ket{B} \\& -\Dot{\Dot{\iota}} \ket{e_2,P_{-2}^2}\ket{g_3,P_0^3}\ket{C} - \ket{e_2,P_{-2}^2}\ket{e_3,P_{-2}^3}\ket{D} \Big] \\& \otimes \prod_{i=A,B} \frac{1}{\sqrt{2}} \big(\ket{0_i} + \ket{1_i} \big).
    \end{split}    
\end{equation}
Here, atoms 1 and 4 will not interact with cavities A and B; therefore, for simplicity, we have denoted the non-interactive atomic states by 
    \begin{align*}    
     \left|A\right> &= \left|g_1,g_4,P_0^1,P_0^4\right> \\
     \left|B\right> &= \left|g_1,e_4,P_0^1,P_{-2}^4\right> \\
     \left|C\right> &= \left|e_1,g_4,P_{-2}^1,P_0^4\right> \\
     \left|D\right> &= \left|e_1,e_4,P_{-2}^1,P_{-2}^4\right>
    \end{align*}
According to our scheme, the undeflected arms of atoms 2 and 3, $\left| g_2,P_0^2\right>$ and $\left|g_3,P_0^3\right>$, will now interact with cavity A, while the deflected arms $\left| e_2,P_{-2}^2\right>$ and $\left|e_3,P_{-2}^3\right>$ interact with cavity B; this is illustrated in Fig. (\ref{fig2}). These interactions are off-resonant Bragg diffractions, governed by the Hamiltonian given in Eq. (\ref{Eq:1}). States of atoms in ground state interacting with cavity A obey similar initial conditions as discussed in section \ref{ABD}. For this case, the governing equations are presented in Eq. (\ref{Eq:3}), Eq. (\ref{Eq:4}) and Eq. (\ref{Eq:5}), and the resulting probability amplitudes are given in Eq. (\ref{Eq:13}) and Eq. (\ref{Eq:14}). However, the governing equations corresponding to Eq. (\ref{Eq:3}), Eq. (\ref{Eq:4}) and Eq. (\ref{Eq:5}) for the split wavepackets of atoms 2 and 3 ($\left| e_2,P_{-2}^2\right>$ and $\left|e_3,P_{-2}^3\right>$), in which the internal states of the atoms are initially in the excited states, will follow from Eq. (\ref{Eq:27}), Eq. (\ref{Eq:28}) and Eq. (\ref{Eq:29}) in Appendix A. In this case, the time-evolved probability amplitudes are also given in Appendix A. This completes the interaction of the respective states of atoms 2 and 3 with cavity A and B, respectively, and the time-evolved state at this stage becomes: 
\begin{equation}
    \begin{split}
        \ket{\psi} =& \frac{1}{4} \bigg[\ket{0_A,0_B} \Big( \ket{g_2,g_3,P_0^2,P_0^3} \ket{A} \\&- \ket{g_2,e_3,P_0^2,P_0^3} \ket{B} - \ket{e_2,g_3,P_0^2,P_0^3} \ket{C} \\& + \ket{e_2,e_3,P_0^2,P_0^3} \ket{D} \Big) \\& + \ket{0_A,1_B} \Big( \ket{g_2,g_3,P_0^2,P_0^3} \ket{A} \\& -\Dot{\iota} \ket{g_2,e_3,P_0^2,P_{-2}^3}\ket{B} +\Dot{\iota} \ket{e_2,g_3,P_{-2}^2,P_0^3} \ket{C} \\&- \ket{e_2,e_3,P_{-2}^2,P_{-2}^3} \ket{D} \Big) \\& + \ket{1_A,0_B} \Big( -\ket{g_2,g_3,P_{-2}^2,P_{-2}^3} \ket{A} \\& + \Dot{\iota} \ket{g_2,e_3,P_{-2}^2,P_0^3} \ket{B} -\Dot{\iota} \ket{e_2,g_3,P_0^2,P_{-2}^3} \ket{C} \\&+ \ket{e_2,e_3,P_0^2,P_0^3} \ket{D} \Big) \\& + \ket{1_A,1_B} \Big( -\ket{g_2,g_3,P_{-2}^2,P_{-2}^3} \ket{A}\\& - \ket{g_2,e_3,P_{-2}^2,P_{-2}^3} \ket{B} - \ket{e_2,g_3,P_{-2}^2,P_{-2}^3} \ket{C} \\&- \ket{e_2,e_3,P_{-2}^2,P_{-2}^3} \ket{D} \Big) \bigg].
    \end{split}
\end{equation}
This equation shows that cavities A and B and both atoms 2 and 3 are mutually sharing information. To swap the information of the cavities only to atoms 2 and 3, an auxiliary atom initially in the ground state ($\ket{g_s}$ and $\ket{g_t}$) passes through each cavity in turn while obeying the resonant interaction with the Hamiltonian $H_R = \hbar \mu \left( \sigma_+ b + \sigma_- b^{\dagger} \right)$ \cite{khan1999quantum}. These auxiliary atoms are detected in one of the possible combinations after passing through the Ramsey zone and traced out from the whole system. Here, we have detected them in their ground states, but the other combinations also yield similar physical results. Now, we are only left with a state in which atoms are correlated with each other, whereas all the cavities are traced out from the state vector. The state of the system becomes:
\begin{equation} \label{Eq:25}
    \begin{split}
    \ket{\psi} = & \frac{1-\Dot{\Dot{\iota}}}{8} \bigg[ \Big( \ket{ g_2,g_3,P_0^2,P_0^3} + \Dot{\Dot{\iota}} \ket{ g_2,g_3,P_{-2}^2,P_{-2}^3} \Big)\bigg]\\& \otimes \ket{ A}  \\& -\frac{1}{8} \bigg[ \Big( \ket{ g_2,e_3,P_0^2,P_0^3} + \ket{ g_2,e_3,P_0^2,P_{-2}^3} \\& - \ket{ g_2,e_3,P_{-2}^2,P_0^3} - \ket{ g_2,e_3,P_{-2}^2,P_{-2}^3} \Big)\bigg]\\& \otimes \ket{B} \\& -\frac{1}{8} \bigg[ \Big( \ket{ e_2,g_3,P_0^2,P_0^3} - \ket{ e_2,g_3,P_0^2,P_{-2}^3} \\& + \ket{ e_2,g_3,P_{-2}^2,P_0^3} - \ket{ e_2,g_3,P_{-2}^2,P_{-2}^3} \Big)\bigg] \\&\otimes \ket{C} \\& + \frac{1-\Dot{\Dot{\iota}}}{8} \bigg[ \Big( \ket{ e_2,e_3,P_{0}^2,P_{0}^3} +\Dot{\iota} \ket{ e_2,e_3,P_{-2}^2,P_{-2}^3} \Big) \bigg] \\& \otimes \ket{ D}.
    \end{split}
\end{equation}
The final step of our scheme is to detect the internal and external states of atoms 2 and 3 after they are passed through a Ramsey zone. Depending on the result of our detection, atoms 1 and 4 may share a different entangled state. The following state represents all the possible entangled states corresponding to the detections on atoms 2 and 3.
\begin{equation}
    \begin{split}
        &\ket{\psi}=\frac{1}{16}\sum_{b=1,3}\sum_{a=1,3}\\&\bigg[\Big ( (1-\Dot{\Dot{\iota}})\ket{A}-\iota^a\ket{B}-\iota^b\ket{C}+(1-\Dot{\Dot{\iota}})\ket{D}\Big)\\&\ket{b_2,a_3,P_0^2,P_0^3}\\& + \Big( (1+\Dot{\Dot{\iota}})\ket{A}+\iota^a\ket{B}+\iota^b\ket{C}+(\Dot{1+\Dot{\iota}})\ket{D} \Big)\\&\ket{b_2,a_3,P_{-2}^2,P_{-2}^3}\\& +\Big( (-\iota^a\ket{B}+\iota^b\ket{C}\Big) \ket{b_2,a_3,P_{0}^2,P_{-2}^3}\\& + \Big( (\iota^a\ket{B}-\iota^b\ket{C} \Big)\ket{b_2,a_3,P_{-2}^2,P_{0}^3}\bigg]. \label{eq24}
    \end{split}
\end{equation}
In Eq. \ref{eq24}, the internal states $\ket{1}$ and $\ket{3}$ refer to the states $\ket{g}$ and $\ket{e}$ of both atoms, respectively. Similarly, we have different possibilities of entangled states of atoms 1 and 4 according to the  detections of atoms 2 and 3. For example, assume that detections on atoms 2 and 4 yields  $\ket{g_2,g_3,P_{0}^2,P_{-2}^3}$, then we are left with the following state:
\begin{equation}
    \begin{split}
        \ket{\psi}= -\ket{B}+\ket{C},
    \end{split}
\end{equation}
which, after expanding the states $\ket{B}$ and $\ket{C}$, becomes:
\begin{equation}
    \begin{split}
        \ket{\psi}= (\ket{e_1,g_4,P_{-2}^1,P_0^4}-\ket{g_1,e_4,P_0^1,P_{-2}^4})/\sqrt{2}. \label{eq26}
    \end{split}
\end{equation}
Eq. \ref{eq26} represents hyperentanglement among the internal and external degrees of freedom of atoms 1 and 4. We note that atoms 1 and 4 do not share a direct interaction with each other in our scheme, and we have successfully swapped hyper-entanglement from atoms 1 and 2 and atoms 3 and 4 to atoms 1 and 4, eventually. 

\section{Delayed-Choice Entanglement Swapping} \label{delayed}
The idea of delayed choice entanglement swapping revolves around the manipulation of entangled particles in a way that their observed properties are seemingly influenced by decisions made at a later time and even at a distant location. Here, we will explore two key variants: bipartite entanglement swapping, which involves the interaction of two entangled pairs, and n-partite entanglement swapping, where the phenomenon extends to multiple entangled particles. These variations shed light on the interconnected nature of quantum systems and the profound influence of observation on the behavior of entangled particles.
\subsection{Bipartite Entanglement}\label{bipar}
Entanglement swapping as described in Section \ref{Ent_swap} can also be performed in a delayed-choice manner \cite{peres2000delayed}, which is more interesting in the context of quantum information, as it highlights the importance of information and information eraser in various quantum protocols. It also explicitly demonstrates the irrelevance of temporal ordering as quantum systems are being processed under unitary dynamics \cite{pan1998experimental}. In our context, we adopt the same procedure as discussed in Section \ref{Ent_swap} to obtain two states of entangled bipartite atomic external momenta, provided for further manipulation as required for delay choice swapping. This pair of entangled states may be expressed in a single mathematical form as follows:
\begin{equation} \label{Eq:27}
\begin{split}
        \ket{\psi^{1,2} (t)} = &\frac{1}{\sqrt{2}} \big[ \ket{0_1,g_1,g_2,P_0^1,P_0^2} \\& + \ket{1_1,g_1,g_2,P_{-2}^1,P_{-2}^2} \big]
\end{split}
\end{equation}
and
\begin{equation} \label{Eq:28}
    \begin{split}
        \ket{\psi^{3,4} (t)} = &\frac{1}{\sqrt{2}} \big[ \ket{0_2,g_3,g_4,P_0^3,P_0^4} \\& + \ket{1_2,g_3,g_4,P_{-2}^3,P_{-2}^4} \big].
    \end{split}
\end{equation}
Here cavity 1 and cavity 2 are still entangled with their respective atoms via their quantized external momenta states: $\ket{P_0^{(j)}}, \ket{P_{-2}^{(j)}}$ with $j=1,2,3,4$. Also, all atoms in free flight are still in their individual ground states $g_{ij}$, again with $j=1,2,3,4$. These control variables are kept unchanged 
so they can be used in future delayed-choice manipulation. 

\begin{figure}[htbp]
       \centering
     \includegraphics[width=8cm]{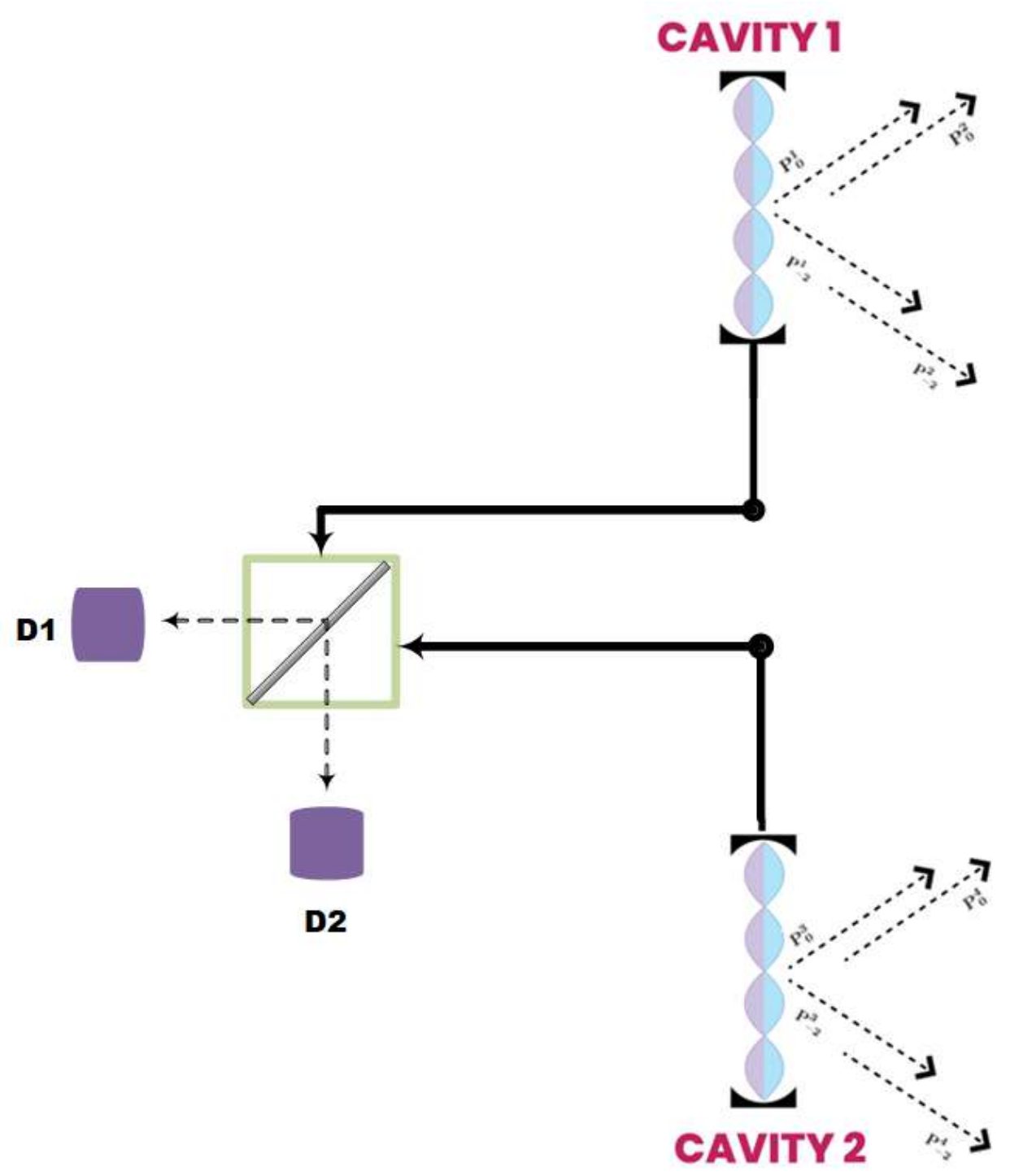}
     \caption{Schematics for implementing Entanglement Swapping: A delayed-choice scenario.}
     \label{fig3}
\end{figure}

In this section, we only demonstrate the basic process of delayed-choice swapping, and therefore we will not discuss the phases involved in various stages. For a symmetric Ramsey Zone (RZ), we take the transformations:
\begin{equation}
    \begin{split}
        \ket{g_A} \rightarrow \frac{1}{\sqrt{2}} \big( \ket{g_{A_j}} + \ket{e_{A_j}} \big)
    \end{split}
\end{equation}
and 
\begin{equation}
    \begin{split}
        \ket{e_A} = \frac{1}{\sqrt{2}} \big( \ket{g_{A_j}} - \ket{e_{A_j}} \big).
    \end{split}
\end{equation}
The subscript $A_j$ designates the two-level atoms that will be used later to erase the quantum information carried by cavity 1 and cavity 2 (in this case, $j=1,2$). Similarly, the symmetric Hadamard Transformations (HT) for atomic quantized external momenta states (atomic momenta beam splitter transformations) may also be expressed as:
\begin{equation}
    \begin{split}
        \ket{P_0^{(j)}} \rightarrow \frac{1}{\sqrt{2}} \big( \ket{P_0^{(j)} + \ket{P_{-2}^{(j)}}} 
    \end{split}
\end{equation}
and
\begin{equation}
    \begin{split}
        \ket{P_{-2}^{(j)}} \rightarrow \frac{1}{\sqrt{2}} \big( \ket{P_0^{(j)} - \ket{P_{-2}^{(j)}}}, 
    \end{split}
\end{equation}
evidently with $j=1,2,3,4$.\\
For the initial state, we tensor product expressions (Eq. \ref{Eq:27}) and (\ref{Eq:28}) to get:
\begin{equation} \label{Eq:33}
    \begin{split}
        &\ket{\psi^{1,2,3,4}} = \ket{\psi^{1,2}} \otimes \ket{\psi^{3,4}} = \\& \frac{1}{\sqrt{2}}.\frac{1}{\sqrt{2}} \Bigg[ \ket{0_1,0_2,g_1,g_2,g_3,g_4,P_0^1,P_0^2,P_0^3,P_0^4} \\&+ \ket{0_1,0_2,g_1,g_2,g_3,g_4,P_0^1,P_0^2,P_{-2}^3,P_{-2}^4} \\&+ \ket{0_1,0_2,g_1,g_2,g_3,g_4,P_{-2}^1,P_{-2}^2,P_0^3,P_0^4} \\&+ \ket{0_1,0_2,g_1,g_2,g_3,g_4,P_{-2}^1,P_{-2}^2,P_{-2}^3,P_{-2}^4} \Bigg].
    \end{split}
\end{equation}
For the intended entanglement swapping, we project both cavity fields simultaneously over a symmetric beam splitter as shown in figure \ref{fig3}.
As illustrated in Fig. \ref{fig3}, this swapping procedure is inherently probabilistic because no photon and two-photon detection events at either of the detectors mark a failure case, which happens half of the time (i.e. with 50 percent probability). Thus, the initial state leading to successful entanglement swapping will be left with
\begin{equation} \label{Eq:34}
    \begin{split}
        &\ket{\psi^{1,2,3,4}} = \\& \frac{1}{\sqrt{2}} \Bigg[ \ket{0_1,1_2,g_1,g_2,g_3,g_4,P_0^1,P_0^2,P_{-2}^3,P_{-2}^4} \\& + \ket{1_1,0_2,g_1,g_2,g_3,g_4,P_{-2}^1,P_{-2}^2,P_{0}^3,P_{0}^4} \Bigg].
    \end{split}
\end{equation}
Projection over the beam splitter is, in fact, the Bell-basis measurement and it effectively erases which-path or which-cavity information carried a priori by the photons. Furthermore, the photons have identical parameters and are simultaneously subjected to the beam splitter and detected at $D_1$ or $D_2$. Thus, after emerging from the beam splitter, the photon subscripts 1 or 2, designating the individual cavities, are no longer relevant to further calculations as they are traced out of the system.

The symmetric beams splitter is one of the most simple quantum devices for the coupling of the inputs to the outputs (optical in and out posts) and is mainly characterised by the following transformations: \cite{gerry2005introductory},
\begin{equation} \label{Eq:35}
    \begin{split}
        \hat{a}_2 = \frac{1}{\sqrt{2}} (\hat{a}_0 + \Dot{\iota} \hat{a}_1), \quad      \hat{a}_3 = \frac{1}{\sqrt{2}} (\Dot{\iota}\hat{a}_0 + \hat{a}_1).
    \end{split}
\end{equation}
Simply stated, an incident single atom, whenever deflected through a beam splitter, symmetric or otherwise, emerges with the phase $\pi/2$, while on the other hand, the transmitted component receives no phase as it passes through without deflection.

Thus the state expressed in Eq. (\ref{Eq:34}) after the applications of the above mentioned beam splitter transformation comes to be,
\begin{equation} \label{Eq:37}
    \begin{split}
        &\ket{\psi^{1,2,3,4}}_{BS} =\\& \frac{1}{\sqrt{2}} \Bigg[ \frac{1}{\sqrt{2}} \big( \ket{g_1,g_2,g_3,g_4,P_0^1,P_0^2,P_{-2}^3,P_{-2}^4} \\&+ \Dot{\iota} \ket{g_1,g_2,g_3,g_4,P_{-2}^1,P_{-2}^2,P_0^3,P_0^4}
        \big) \otimes \ket{1_{D_1}} \\&+ \frac{1}{\sqrt{2}} \big( \Dot{\iota} \ket{g_1,g_2,g_3,g_4,P_0^1,P_0^2,P_{-2}^3,P_{-2}^4} \\& + \ket{g_1,g_2,g_3,g_4,P_{-2}^1,P_{-2}^2,P_0^3,P_0^4} \big) \otimes \ket{1_{D_2}} \Bigg].
    \end{split}
\end{equation}
In Eq. (\ref{Eq:37}), we traced out the common vacuum state. This completes the simple entanglement swapping protocol in the external quantized momenta states. 

At the receiver's side, the receiver can locally transform this four-partite entangled momenta state into a hyperentangled state (i.e., state entangled in two degrees of freedom: internal and external states) by exposing a pair of momenta components of any of the two atom pairs  (atom 1, atom 2) or (atom 3, atom 4) to a classical Ramsey field in a delayed-choice manner. This type of interaction is governed by the resonant semi-classical interaction Hamiltonian $\nu = -h\Omega_R (\sigma_+ + \sigma_-)/2$ with $\Omega_R$ being the resonant Rabi frequency and $\sigma+(\sigma_-)$ is the atomic raising/lowering operator \cite{zubairy1997quantum}. 

From expression (\ref{Eq:37}), the entangled state engineered through the swapping when the photon is detected at the detector $D_1$, is given by:
\begin{equation} \label{Eq:38}
    \begin{split}
        \ket{\psi^{1,2,3,4}} = & \frac{1}{\sqrt{2}} \Bigg[\ket{g_1,g_2,g_3,g_4,P_{0}^1,P_{0}^2,P_{-2}^3,P_{-2}^4} \\& + \Dot{\iota} \ket{g_1,g_2,g_3,g_4,P_{-2}^1,P_{-2}^2,P_0^3,P_0^4} \Bigg],
    \end{split}
\end{equation}
Exposing the atomic momenta components $\ket{P_0^1,P_0^2}$ of atom 1 and atom 2 as well as $\ket{P_{-2}^3,P_{-2}^4}$ of atom 3 and atom 4 to a classical laser field with interaction time corresponding to $\pi$ Rabi pulse causes the transformations of the internal state of the atom,
\begin{equation}
    \begin{split}
        \ket{g} \rightarrow \Dot{\iota} \ket{e},\quad  \ket{e} \rightarrow \Dot{\iota} \ket{g}.
    \end{split}
\end{equation}
Hence, expression (\ref{Eq:38}), under laser-atom interaction for the specified time, transforms to the following four-partite hyperentangled state, i.e. the state entangled in the internal as well as the external momenta states in accordance with the systematic depicted in figure (\ref{fig3}). Finally, we obtain the delayed-choice engineered hyperentangled state with the receiver opting for atom-laser field interactions with appropriate later time. The desired state, will, therefore be:
\begin{equation}
    \begin{split}
        \ket{\psi^{1,2,3,4}}^{hypent} = & \frac{1}{\sqrt{2}} \Bigg[\ket{e_1,e_2,e_3,e_4,P_{0}^1,P_{0}^2,P_{-2}^3,P_{-2}^4} \\&+ \Dot{\iota} \ket{g_1,g_2,g_3,g_4,P_{-2}^1,P_{-2}^2,P_0^3,P_0^4} \Bigg].
    \end{split}
\end{equation}
This is because four atoms have successively gone through atom-field interactions leading to the change of the respective ground states into the corresponding excited state, each producing a local phase of $\Dot{\iota}$ and collectively $\Dot{\iota}^4=1$.

\subsection{n-partite Entanglement}
We can generalize bipartite entanglement swapping and its delayed-choice based conversion into n-partite hyperentanglement swapping. Experimentally, it has been well demonstrated that a stream of thousands of atoms can interact with a cavity field consecutively before the appreciable inset of decoherence \cite{haroche2006exploring,islam2008generation,khosa2004measurement}. This is mainly due to the availability of high-Q superconducting cavities with useful life as high as a fraction of a second \cite{deleglise2008reconstruction,kuhr2007ultrahigh}. Thus, instead of Bragg diffracting only a single pair of atoms from each cavity, we may diffract $n/2$ atoms, where $n \geq 500$, from each cavity \cite{ali2022generation}.  Following the procedure described in section \ref{bipar}, the pair of the generated entangled states of the $n/2$-partite may be expressed as follows.
\begin{equation} \label{Eq:40}
\begin{split}
    \ket{\psi}^{n/2}_{C_1} &= \frac{1}{\sqrt{2}} \Bigg[ \ket{0_1} \otimes \prod_{i=1}^{n/2} \bigg( \ket{g_i,P_0^{(i)}} \bigg) \\
    &\quad + \ket{1_1} \otimes \prod_{i=1}^{n/2} \bigg( \ket{g_i,P_{-2}^{(i)}} \bigg) \Bigg]
\end{split}
\end{equation}
and,
\begin{equation} \label{Eq:41}
    \begin{split}
        \ket{\psi}^{n/2}_{C_2} = & \frac{1}{\sqrt{2}} \Bigg[ \ket{0_2} \otimes \Pi^{n}_{j=\frac{n}{2}+1} \bigg( \ket{g_j,P_0^{(j)}} \bigg) \\& + \ket{1_2} \otimes \Pi^{n}_{j=\frac{n}{2}+1} \bigg( \ket{g_j,P_{-2}^{(j)}} \bigg) \Bigg].
    \end{split}
\end{equation}
Again, by employing the same procedure cited earlier, i.e. by projecting cavity fields simultaneously over a symmetric beam splitter, we can engineer an n-partite entangled state through swapping. This procedure yields the state, 
\begin{equation}
    \begin{split}
         & \ket{\psi}^n =\\& \frac{1}{\sqrt{2}} \Bigg[ \bigg[\Pi^{n/2}_{j=1} \bigg(\ket{g_j,P_0^{(j)}} \bigg) \otimes \Pi^{n}_{k=\frac{n}{2}+1} \bigg(\ket{g_k,P_{-2}^{(k)}} \bigg) \bigg] \\&+ \Dot{\iota} \bigg[ \Pi^{n/2}_{j=1} \bigg(\ket{g_j,P_{-2}^{(j)}} \bigg)\otimes \Pi^{n}_{k=\frac{n}{2}+1} \bigg(\ket{g_k,P_{0}^{(k)}} \bigg) \bigg] \Bigg]
    \end{split}
\end{equation}
This state, once in the access of the receiver, can be transformed into the hyperentangled state in a delayed-choice manner by passing the specifically selected momentum components through classical Ramsey fields, where the split wave packet of each atom is excited from its respective ground state following the procedure described at length in the previous section. The procedure will duly yield the following hyperentangled state,
\begin{equation}
    \begin{split}
         & \ket{\psi}^{n-hypent} =\\& \frac{1}{\sqrt{2}} \Bigg[ \bigg[ (\Dot{\iota}^n \Pi^{n/2}_{j=1} \bigg(\ket{e_j,P_0^{(j)}} \bigg) \otimes \Pi^{n}_{k=\frac{n}{2}+1} \bigg(\ket{e_k,P_{-2}^{(k)}} \bigg) \bigg] \\&+ \Dot{\iota} \bigg[ \Pi^{n/2}_{j=1} \bigg(\ket{g_j,P_{-2}^{(j)}} \bigg)\otimes \Pi^{n}_{k=\frac{n}{2}+1} \bigg(\ket{g_k,P_{0}^{(k)}} \bigg) \bigg] \Bigg]
    \end{split}
\end{equation}
or,
\begin{equation}
    \begin{split}
         & \ket{\psi}^{n-hypent} =\\& \frac{1}{\sqrt{2}} \Bigg[ \bigg[\Dot{\iota}^{n-1} \Pi^{n/2}_{j=1} \bigg(\ket{e_j,P_0^{(j)}} \bigg) \otimes \Pi^{n}_{k=\frac{n}{2}+1} \bigg(\ket{e_k,P_{-2}^{(k)}} \bigg) \bigg] \\&+ \bigg[ \Pi^{n/2}_{j=1} \bigg(\ket{g_j,P_{-2}^{(j)}} \bigg)\otimes \Pi^{n}_{k=\frac{n}{2}+1} \bigg(\ket{g_k,P_{0}^{(k)}} \bigg) \bigg] \Bigg].
    \end{split}
\end{equation}
The procedure adopted for the generation of entanglement, swapping and the subsequent delayed-choice transformation into the n-partite hyperentanglement engineering, is universal and hence can be utilized for diverse entanglement morphology including cluster and graph states \cite{markham2008graph,lu2007experimental}. Thus, the protocol presented here may contribute to the realization of complex quantum entangled networks and other applications requiring information distribution over complex quantum structures, such as many-body systems, and simulation of biological systems. \cite{santiago2020quantum}.

\section{Experimental Feasibility} \label{experiment}
Quantum entanglement, along with its characteristic traits, is the most prominent resource in the field of quantum information \cite{einstein1935can,bell1964einstein,aspect1982experimental,nielsen2010quantum, megidish2013entanglement, raimond2001manipulating}. Similarly, entanglement swapping is a central process that has been used to clarify a foundational issue of entanglement engineering, i.e. the demonstration of entanglement between two parties without any direct temporal or spatial interaction of the parties involved. This was carried out through Bell-basis measurements on two independent but entangled pairs, which consequently placed information (and information eraser) as the core theme in quantum information theory \cite{peres2000delayed}. 
The advent of hyperentanglement, and its utilization in quantum information, further fortified the quantum information domain, as it significantly enhances the channel capacity without invoking any further physical resources, which minimizes resource-related quantitative decoherence. Cavity QED-based atom-field systems have been historically the pioneering gadgets for efficient and meaningful demonstrations and experimental implementations of many quantum information proposals. 
\cite{haroche2006exploring,haroche2013nobel}. However, the main problem with such multi-partite quantum states is the decoherence. This is because many quantum information experiments generally require precise interactions under a scenario that is not ideally closed, eventually leading to decoherence. 


Our proposed hyperentanglement swapping fundamentally utilizes external momenta states of the neutral atoms, and such states are well known to be resistant to decoherence under what is generally termed quantum Darwinism \cite{ball2008quantum}. For state engineering, we rely on the off-resonant Bragg diffraction of the neutral atoms from the cavity fields. This subsequently generates the desired states in momentum space without exchanging any real photons between the atom and the field. Therefore, in our work, the pathway for the decoherence emanating from spontaneous emission is also effectively closed. Currently, we can engineer cavities with lifetimes close to a fraction of a second \cite{rempe1990observation,gerry1996proposal}, which successfully tackles decoherence. This means that several atoms can interact with a quantized cavity field without raising substantial threats to the coherence of the system \cite{deleglise2008reconstruction}. It is also reasonable to treat the system as closed because the average interaction time between an atom and a cavity field is in the range of microseconds. Over the years, numerous experimental validations for atomic Bragg diffraction have been performed, with even $8^{th}$ order atomic Bragg diffraction generating experimentally fruitful results \cite{durr1996pendellosung,vernooy1998high,durr1998origin}. Mach-Zehnder-Bragg interferometry has also been applied experimentally with results involving splitting of momenta of 102 photons \cite{chiow2011102}. Furthermore, the transmission of the resulting momentum of around 112 $\hbar k$ has also been experimentally observed with high visibility of the fringes \cite{plotkin2018three,manning2015wheeler}. Therefore, in light of these experimental demonstrations, it may be safely assumed that our proposed scheme can be realized experimentally. 

Moreover, the proposed scheme holds practical grounds due to the fact that high-fidelity systems have previously been proposed and executed efficiently in the microwave Bragg regime. In the optical regime, experiments have also shown generally good and reliable Bragg diffraction of atoms. In terms of experimental efficiency, variational errors in interaction time can be avoided by utilizing ultra-cold atoms trapped by magneto-optics with a negligible spread in velocity \cite{durr1996pendellosung,durr1998origin,kunze1996bragg,giltner1995theoretical}. 
To be more specific, the atomic Bragg diffraction of $^{85}$Rb atoms by $780$ nm optical laser beams \cite{rempe1990observation,gerry1996proposal,durr1998origin} provides a glaring experimental realization where atoms with mass $M = 85$ amu are exploited with recoil frequency $\omega_r = 2.4$ x $10^4$ rad/s and Rabi frequency $\Omega = 2\pi$ x $16.4$ MHz. The parameters related to the cavity include detuning of $1$ GHz and finesse $4.4$ x $10^5$. As an alternative, $^{4}$He atoms with $M = 4$ amu, $\omega_r = 1.06$ MHz, $\lambda= 543.5$ nm, $\Delta = 6.28$ GHz and the effective Rabi frequency $\frac{\mu^2}{4\Delta} = 120 $ KHz can also be used for an interaction time of $13 \mu$s the finesse of the cavities, i.e $F = 7.85$ x $10^6$, leading to the lifetimes up to milliseconds \cite{khosa2004measurement,hood2001characterization}. 
Consequently, these atomic and cavity parameters maintain the conditions for off-resonant atomic Bragg diffraction, i.e $\Delta >> \omega_r$ and $\omega_r + \Delta = \mu \sqrt{n}/2$, 
where the interaction time is kept far lower than the lifetime of the cavity (around $0.5 \mu$s) \cite{durr1996pendellosung,munstermann1999dynamics,munstermann1999single,puppe2004single}.

\section{Conclusion}\label{conclusion}
In this paper, we proposed a framework for entanglement swapping of hyperentanglement between two hyperentangled pairs. One atom from each pair is manipulated while we do not interfere with the two entangled atoms. Passing the undeflected arms from Cavity A and deflected arms from Cavity B, we produce a hyper-correlation between the two atoms. Finally, detection of these atom's internal and external states after passing them through Ramsey zone, swaps the entanglement on to the atoms which passed without interactions initially. Such a protocol, if implemented experimentally, will ensure the safe decoherence-free and high-capacity information transfer over two quantum nodes or quantum information processing units that were independent and uncorrelated initially.
\bibliographystyle{unsrt}
\bibliography{main}
\section*{Appendix A:Interaction between the atom in excited state and the cavity in superposition state}
If the atom is in the excited state then the initial state vector will be,
\begin{equation}
    \ket{\psi(t=0)} = \ket{g,P_0} \otimes (\ket{0}+\ket{1})/\sqrt{2}
\end{equation}

And the proposed state vector can be written as,
\begin{equation}
\begin{split}
    \left| \psi (t) \right> = &e^{-\Dot{\iota} \left( \frac{P_0^2}{2M \hbar}-\frac{\Delta}{2}\right)t} \sum_{l=-\infty}^{\infty} \left[ C_{e,0}^{P_l}(t)\left|e,0,P_l\right>+\right.\\&\left.  C_{e,1}^{P_l}(t) \left|e,1,P_l\right> +  C_{g,1}^{P_l}(t) \left|g,1,P_l\right> \right].
\end{split}
\end{equation}

Using the same governing Hamiltonian as presented in Eq. (\ref{Eq:1}), we solve the Schrodinger equation and after taking projections of probability amplitudes, we arrive at the following coupled differential equations,
\begin{equation} \label{Eq:47}
\begin{split}
        \Dot{\iota} \frac{\partial}{\partial t} C_{e,0}^{P_l}(t) =& \left( \frac{l(l_0 +l)\hbar k^2}{2M} \right) C_{e,0}^{P_l}(t) + \\&\frac{\mu}{2} \left[ C_{g,1}^{P_{l+1}}(t) + C_{g,1}^{P_{l-1}}(t) \right],
\end{split}
\end{equation} 

\begin{equation} \label{Eq:48}
    \Dot{\iota} \frac{\partial}{\partial t} C_{e,1}^{P_{l}}(t) = \left( \frac{l(l_0 +l)\hbar k^2}{2M} \right) C_{e,1}^{P_{l}}(t),
\end{equation} 

\begin{equation} \label{Eq:29}
\begin{split}
    \Dot{\iota} \frac{\partial}{\partial t} C_{g,1}^{P_l}(t) =& \Delta C_{g,1}^{P_l}(t) +\\& \frac{\mu}{2} \left[ C_{e,0}^{P_{l+1}}(t) + C_{e,0}^{P_{l-1}}(t) \right].
\end{split}
\end{equation}

Putting in the values of $l$ which runs from $-3$ to $1$, we get,
\\\\
For l=1, \begin{equation}
\Dot{\iota} \frac{\partial}{\partial t}C_{g,1}^{P_{1}}(t)=  \Delta C_{g,1}^{P_{1}} (t) + \frac{\mu}{2} \left(C_{e,0}^{P_{2}}(t) + C_{e,0}^{P_{0}}(t) \right),
\end{equation} 
\\
For l=0, \begin{equation}
    \Dot{\iota} \frac{\partial}{\partial t}C_{e,0}^{P_0}(t) = 
    \frac{\mu}{2} \left( C_{g,1}^{P_{1}} (t) +C_{g,1}^{P_{-1}} (t) \right),
\end{equation} 
\\
For l=-1,
\begin{equation}
    \Dot{\iota} \frac{\partial}{\partial t}C_{g,1}^{P_{-1}}(t)=  \Delta C_{g,1}^{P_{-1}} (t) + \frac{\mu}{2} \left(C_{e,0}^{P_{0}}(t) + C_{e,0}^{P_{-2}}(t) \right), 
\end{equation} 
\\
For l=-2,
\begin{equation}
     \Dot{\iota} \frac{\partial}{\partial t}C_{e,0}^{P_-2}(t) = \frac{\mu}{2} \left( C_{g,1}^{P_{-1}} (t) +C_{g,1}^{P_{-3}} (t) \right),
\end{equation} 
\\
For l=-3,
\begin{equation}
    \Dot{\iota} \frac{\partial}{\partial t}C_{g,1}^{P_{-3}}(t)= \Delta C_{g,1}^{P_{-3}} (t) + \frac{\mu}{2} \left(C_{e,0}^{P_{-2}}(t) + C_{e,0}^{P_{-4}}(t) \right).
\end{equation}

Applying adiabatic approximation and back substitutions, we reach,
\begin{equation}
    \frac{\partial}{\partial t} C_{e,0}^{P_0}(t) = \frac{\Dot{\iota} \mu^2}{4 \Delta} \left[ 2C_{e,0}^{P_0}(t) + C_{e,0}^{P_-2}(t) \right],
\end{equation} 

\begin{equation}
   \frac{\partial}{\partial t}  C_{e,0}^{P_{-2}}(t) = \frac{\Dot{\iota} \mu^2 }{4 \Delta} \left[ C_{e,0}^{P_-2}(t) +  C_{e,0}^{P_0}(t) \right].
\end{equation}

Solving the above coupled differential equation, we finally arrive at the governing equations for this system,

\begin{equation}
\begin{split}
    C_{e,0}^{P_0}(t) =& e^{\frac{2\Dot{\iota} \mu^2 t}{4\Delta}} \left[  C_{e,0}^{P_0}(0) \cos\left(\frac{\mu^2}{4\Delta} t\right) \right.\\&\left. + \Dot{\iota} C_{e,0}^{P_-2}(0)\sin\left(\frac{\mu^2}{4\Delta} t\right) \right],
\end{split}
\end{equation}

\begin{equation}
\begin{split}
    C_{e,0}^{P_-2}(t) = &e^{\frac{2\Dot{\iota} \mu^2 t}{4\Delta}} \left[  C_{e,0}^{P_-2}(0) \cos\left(\frac{\mu^2}{4\Delta} t\right) \right.\\&\left. + \Dot{\iota} C_{e,0}^{P_0}(0)\sin\left(\frac{\mu^2}{4\Delta} t\right) \right].
\end{split}
\end{equation}

With initial condition, $C_{e,0}^{P_{-2}}(0) = 1/\sqrt{2}$ and setting interaction time as $t = 2\pi\Delta/\mu^2$,

\begin{equation}
    \left|\psi \right>_1 = \frac{1}{\sqrt{2}} \left[ \left| 1,e,P_{-2} \right> -\Dot{\iota} \left| 0,e,P_0 \right> \right].
\end{equation}
\end{document}